\documentstyle[12pt]{article}
\def\be {\begin{equation}}
\def\ee {\end{equation}}
\def\ba {\begin{eqnarray}}
\def\ea {\end{eqnarray}}

\def\bi {\begin{itemize}}
\def\ei {\end{itemize}}
\begin{document}
\def\bea{\begin{eqnarray}}
\def\eea{\end{eqnarray}}
\title{\bf  {The Cosmological Dynamics of Interacting  Holographic Dark Energy Model}}
\author{{M. R. Setare$^{1}$\thanks{%
E-mail: rezakord@ipm.ir} , Elias C. Vagenas$^{2,3}$ \thanks{%
E-mail: evagenas@phys.uoa.gr} }\\
\newline
{\it $^1$ Department of Science, Payame Noor University, Bijar, Iran}
\\
{\it $^2$ Nuclear and Particle Physics Section,
 Physics Department,}\\
 {\it  University of Athens,
 GR-15771, Athens, Greece}
 \\
{\it $^3$
 Research Center for Astronomy \& Applied Mathematics,}\\
 {\it Academy of Athens,
 Soranou Efessiou 4,
 GR-11527, Athens, Greece}}

\date{\small{}}

\maketitle
\begin{abstract}
Motivated by the recent observations for the cosmic acceleration and the suitable evolution of the Universe provided
an interaction (decay of dark energy to matter) is incorporated in a cosmological model, we study the
cosmological evolution of the Interacting Holographic Dark Energy scenario. Critical points are derived and their corresponding
cosmological models are presented. The dynamical character of these models is revealed.
 \end{abstract}
% \begin{document}

\newpage
% \vspace*{10mm}

\section{Introduction}
Recent observations from type Ia supernovae \cite{SN} associated
with Large Scale Structure \cite{LSS} and Cosmic Microwave
Background anisotropies \cite{CMB} have provided main evidence for
the cosmic acceleration. The combined analysis of cosmological
observations suggests that the universe consists of about $70\%$
dark energy, $30\%$ dust matter (cold dark matter plus baryons),
and negligible radiation. Although the nature and origin of dark
energy are unknown, we still can propose some candidates to
describe it, namely  since we do not know where this dark energy comes from,
and how to compute it from the first principles, we search for
phenomenological models. Hopefully, astronomical observations will select
one of them.
%%%%%%%%%%%%%%%%%%%%%%%%%%%%
The most obvious theoretical candidate of dark energy is the
cosmological constant $\lambda$ (or vacuum energy)
\cite{Einstein:1917,cc} which has the equation of state $w=-1$.
However, as is well known, there are two difficulties that arise
from the cosmological constant scenario, namely the two famous
cosmological constant problems --- the ``fine-tuning'' problem and
the ``cosmic coincidence'' problem \cite{coincidence}. An
alternative proposal for dark energy is the dynamical dark energy
scenario. The dynamical dark energy proposal is often realized by
some scalar field mechanism which suggests that the energy form
with negative pressure is provided by a scalar field evolving down
a proper potential. So far, a large class of scalar-field dark
energy models have been studied, including quintessence
\cite{quintessence}, K-essence \cite{kessence}, tachyon
\cite{tachyon}, phantom \cite{phantom}, ghost condensate
\cite{ghost2} and quintom \cite{quintom}, and so forth. But we
should note that the mainstream viewpoint regards the scalar field
dark energy models as an effective description of an underlying
theory of dark energy. In addition, other proposals on dark energy
include interacting dark energy models \cite{intde}, braneworld
models \cite{brane}, Chaplygin gas models \cite{cg},
and many others.\\
%%%%%%%%%%%%%%%%%%%%%%%%%%%%%%%%%%%%%%%%%%%%%%%%%%%%%%%
Currently, an interesting attempt for probing the nature of dark
energy within the framework of quantum gravity
(and thus compute it from first principles)
 is the so-called ``Holographic Dark Energy'' (HDE) proposal
\cite{Cohen:1998zx,Horava:2000tb,Hsu:2004ri,Li:2004rb}. It is well
known that the holographic principle is an important result of the
recent researches for exploring the quantum gravity (or string
theory) \cite{holoprin}.  The Holographic Dark Energy model has
been tested and constrained by various astronomical observations
\cite{Huang:2004wt,obs1,obs2,obs3} as well as by the Anthropic
Principle \cite{Huang:2004mx}. Furthermore, the Holographic Dark
Energy model has been extended to include the spatial curvature
contribution, i.e. the Holographic Dark Energy model in non-flat
space \cite{nonflat}. For other
extensive studies, see e.g. \cite{holoext}\footnote{A very recent development
 is the idea of bulk holographic dark energy.
In this proposal, holographic dark energy is accommodated in the
framework of braneworld cosmology \cite{Saridakis:2007cy}.}.\\
%%%%%%%%%%%%%%%%%%%%%%%%%%%%%%%%
It is known that the coincidence or, ``why now" problem is easily
solved in some models of holographic dark energy based on this fundamental assumption
that matter and holographic dark energy do not conserve separately,
but the matter energy density decays into the holographic energy
density \cite{interac,Amendola:2000uh}. In fact a suitable evolution of the Universe
is obtained when, in addition to the holographic dark energy, an
interaction (decay of
dark energy to matter) is assumed.\\
The remainder of the paper is organized as follows. In Section 2,  we present the Holographic Dark Energy model where the interaction is included.
Since the fractional contributions of the curvature ($\Omega_k$), the ordinary matter ($\Omega_m$),
and the dark energy ($\Omega_{X}$) do not form a compact state space, we introduce  a new set of
dimensionless variables which form such a compact state space. The evolution equations of these variables are derived
and in order to study their dynamical character the critical points of this system are obtained.
Cosmological models that correspond to the aforementioned critical points are presented and their dynamical character is given.
%
%
%%%%%%%%%%%%%%%%%%%%%%%%%%%%%%%%%%%%%%%%%%%%%%
At this point, it is worth noting that one of the properties that
should characterize a cosmological solution is that the present
universe is  a global attractor, i.e. all the possible initial
conditions lead to the observed percentages of dark energy and
dark matter; once reached, they remain fixed forever
\cite{Amendola:2000uh}. This is the reason that here we will focus
on the stability of the cosmological models that correspond to
critical points and more importantly, these cosmological models
are attractors.
%%%%%%%%%%%%%%%%%%%%%%%%%%%%%%%%%%%%%%%%%%%%%%
%
%
Finally, Section 3 is devoted to a brief summary of results and concluding remarks.
%%%%%%%%%%%%%%%%%%%%%%%%%%%%%%%%%%%%%%%%%%%%%%%%%%%%%%%%%%%%%%%%%%%%%%%%%%%%%%%%%%
%%%%%%%%%%%%%%%%%%%%%%%%%%%%%%%%%%%%%%%%%%%%%%%%%%%%%%%%%%%%%%%%%%%%%%%%%%%%%%%%
\section{Stability of Interacting Holographic Dark Energy model solutions}
In this section we consider the Holographic Dark Energy model when
there is an interaction between the holographic energy density
$\rho_{X}$ and  Cold Dark Matter (CDM) which has $w_{m}=0$.
%
%
%%%%%%%%%%%%%%%%%%%%%%%%%%%%%%%%%%
Since we know neither the nature of dark energy nor the nature of dark
matter, a microphysical interaction model is not available either.
However, pressureless dark matter in interaction with
holographic dark energy is more than just another model to describe an
accelerated expansion of the universe. It provides a unifying view of  different models
which are viewed as different realizations of the Interacting Holographic
Dark Energy Model at the perturbative level \cite{Zimdahl:2007ne}.
%%%%%%%%%%%%%%%%%%%%%%%%%%%%%%%%%%%%5
%
%
The continuity equations for dark energy and CDM are
\begin{eqnarray}
\label{2eq1}&& \dot{\rho}_{\rm X}+3H(1+w_{\rm X})\rho_{\rm X} =-Q, \\
\label{2eq2}&& \dot{\rho}_{\rm m}+3H\rho_{\rm m}=Q\hspace{1ex}.
\end{eqnarray}
The interaction is given by the quantity $Q=\Gamma
\rho_{X}$ with $\Gamma$ to have dimensions of pressure. This is a decaying of the holographic energy
component into CDM with the decay rate $\Gamma$. Taking a ratio
of two energy densities as $r=\rho_{\rm m}/\rho_{\rm X}$,
the above equations lead to
\begin{equation}
\label{2eq3} \dot{r}=3Hr\Big[w_{\rm X}+
\frac{1+r}{r}\frac{\Gamma}{3H}\Big]\hspace{1ex}.
\end{equation}
%
%
%%%%%%%%%%%%%%%%%%%%%%%%%%%%%%%%%%%%%%%%%
It should be stressed that the decay rate has to be positive, i.e. $\Gamma>0$,
so as to interpret a transfer from the dark energy component to the matter component. Moreover,
it is obvious from equation (\ref{2eq3}) that any restriction on the parameter of the equation of state for
the holographic dark energy, i.e. $w_{\rm X}$, will set constraints on $\Gamma>0$ \cite{Pavon:2005kr}.
%%%%%%%%%%%%%%%%%%%%%%%%%%%%%%%%%%%%%%%%%%
%
%
\par\noindent
 Following \cite{Kim:2005at},
if we define
\begin{eqnarray}
\label{eff}
w_X ^{\rm eff}=w_X+{{\Gamma}\over {3H}}\;, \qquad w_m
^{\rm eff}=-{1\over r}{{\Gamma}\over {3H}}\; ,
\end{eqnarray}
then the continuity equations can be written in their standard
form
\begin{equation}
\dot{\rho}_X + 3H(1+w_X^{\rm eff})\rho_\Lambda =
0\;,\label{definew1}
\end{equation}
\begin{equation}
\dot{\rho}_m + 3H(1+w_m^{\rm eff})\rho_m = 0\;
\label{definew2}
\hspace{1ex}.
\end{equation}
We consider the non-flat Friedmann-Robertson-Walker universe with
line element
 \be
 \label{metr}
ds^{2}=-dt^{2}+a^{2}(t)(\frac{dr^2}{1-kr^2}+r^2d\Omega^{2})
 \ee
where $k$ denotes the curvature of space with k=0,1,-1 for flat,
closed and open universe respectively. A closed universe with a
small positive curvature ($\Omega_k\sim 0.01$) is compatible with
observations \cite{ {wmap}, {ws}}. We employ the Friedmann equation
to relate the curvature of the universe to the energy density. The
first Friedmann equation is given by
\begin{equation}
\label{2eq7} H^2+\frac{k}{a^2}=\frac{1}{3M^2_p}\Big[
 \rho_{\rm X}+\rho_{\rm m}\Big]\hspace{1ex}.
\end{equation}
%%%%%%%%%%%%%%%%%%%%%%
We define the dimensionless fractional contributions of CDM, holographic dark energy, and
curvature as follows
\begin{equation}
\label{2eq9}
\Omega_{\rm
m}=\frac{\rho_{m}}{\rho_{cr}}=\frac{ \rho_{\rm
m}}{3M_p^2H^2}\hspace{1ex},\hspace{1cm}\Omega_{\rm
X}=\frac{\rho_{X}}{\rho_{cr}}=\frac{ \rho_{\rm
X}}{3M^2_pH^2}\hspace{1ex},\hspace{1cm}\Omega_{k}=\frac{k}{a^2H^2}
\hspace{1ex}.
\end{equation}
%%%%%%%%%%%%%%%%%%%%%%%%%%%%%%%%%%%%%%%%%%%%%%%%
%%%%%%%%%%%%%%%%%%%%%%%%%%%%%%%%%%%%%%%%%%%%%%%%
%%%%%%%%%%%%%%%%%%%%%%%%%%%%%%%%%%%%%%%%%%%%%%%%
Using the results of \cite{set1}, the effective parameters of the equation of states given in equation
(\ref{eff}) are now expressed in terms of the above-mentioned  dimensionless parameters as
\be
\label{we1}
w_{\rm
X}^{eff}=-\frac{1}{3}-\frac{2\sqrt{\Omega_{\rm
X}-c^2\Omega_{k}}}{3c}
\hspace{1ex},
\ee
\be
\label{we2}w_{m}^{eff}=\frac{-b^2(1+\Omega_{k})}{\Omega_{\rm m}}
\hspace{1ex}.
\ee
%%%%%%
At this point it should be noted that for the computation of the effective parameters as given
above, the decay rate is chosen to be
\be
\Gamma= 3 b^{2} \left(1+r\right) H
\ee
where  $b$ is a dimensionless coupling constant, and the holographic dark energy density is of the form
\be
\rho_{X}=3 c^2 M^{2}_{p}L^{-2}
\ee
where $c$ is a positive constant which can not be less than $1$. The IR cut-off parameter $L$ is chosen to
be the radius of the event horizon as measured on the sphere of the horizon and is defined as
\be
L=a r(t)
\ee
where $r(t)$ is obtained from
\be
\int^{r(t)}_{0}\frac{dr}{\sqrt{1-kr^2}}=\frac{R_h}{a}
\hspace{1ex}
\ee
which yields
\be
r(t)=\frac{1}{\sqrt{k}}\sin y
\ee
with $y=\sqrt{k}R_{h}/a$ and $R_h$ is the radius of the event horizon measured in the $r$ direction
\be
R_{h}=a\int^{\infty}_{t}\frac{dt}{a}
\hspace{1ex}.
\ee
%%%%%%%%%%%%%%%%%%%%%%%%%%%%%%%%%
\par\noindent
Now we describe solutions of the model under investigation and
also determine their stability. To do this, we use a set of
convenient phase-space variables similar to those introduced
in \cite{CS,GE}. We introduce the notation similar
to those of \cite{CS}, and work in the $3$-dimensional
$\Omega$-space $(\Omega'_{k},\Omega_{m},\Omega_{\rm X})$ where $\Omega'_{k}=-\Omega_{k}$.
We can now rewrite the first Friedmann equation as
\begin{equation}
\label{2eq10}
\Omega_{\rm m}+\Omega_{\rm
\Lambda}+\Omega'_{k}=1\hspace{1ex}.
\end{equation}
It is obvious that when the  universe is flat ($k=0$) or open ($k=-1$),
all terms in equation (\ref{2eq10}) that are put together to give unity
are nonnegative and take values in $[0,1]$.
Nonetheless all of them are singular in the case of $H=0$.
For the case of closed universe ($k=1$) which is of our interest,
the terms in equation (\ref{2eq10}) are not anymore all nonnegative.
Therefore, the state space defined by the variables
$(\Omega'_{k},\Omega_{m},\Omega_{\rm X})$ is no longer compact
(because now $\Omega'_{k}< 0$ for $k=1$). However, we can employ
another set of variables describing a compact state space. Therefore,
in lieu of using the Hubble function $H$, we introduce the following
quantity
\begin{equation}
D=\sqrt{H^2+\frac{k}{a^2}}\hspace{1ex} \label{ppp}
\end{equation}
and thus we define the following dimensionless variables
\begin{equation}
Z=\frac{H}{D}\hspace{1ex},\label{Z}
\end{equation}
\begin{equation}
\tilde{\Omega}_{m}=\frac{ \rho_{\rm m}}{3M_p^2D^2}\hspace{1ex},\label{m}
\end{equation}
\begin{equation}
\tilde{\Omega}_{X}=\frac{ \rho_{\rm
X}}{3M^2_pD^2}\hspace{1ex}.\label{Lambda}
\end{equation}
From these definitions it is evident that now the case $H = 0$ is included.
Moreover, the Friedmann equation takes the following form
\begin{equation}
\tilde{\Omega}_{m}+\tilde{\Omega}_{X} =1\hspace{1ex}
\label{const1}
\end{equation}
which, together with the fact that $-1\leq Z \leq 1$ since the Hubble  parameter
can be positive ($H>0$) for an expanding cosmological model or negative
($H<0$) for a contracting model,
implies that the state space defined by the new
variables is indeed compact. Introducing the primed time derivative
\begin{equation}
' = \frac{1}{D} \frac{d}{dt}
\hspace{1ex}
\end{equation}
one obtains the system of first-order differential equations
\be
\label{deq}
D'=-Z^{3}D\left(q+\frac{1}{Z^2}\right)\hspace{1ex} ,
\ee
\be
\label{zeq}
Z'=Z^2\left[-1-q+Z^{2}\left(q+\frac{1}{Z^2}\right)\right] \hspace{1ex} ,
\ee
\be
\label{omm}
\tilde{\Omega}'_{m}=\tilde{\Omega}_{m}Z\left[-3\left(1+w_m ^{\rm
eff}\right)+2Z^2 \left(q+\frac{1}{Z^2}\right)\right]
\hspace{1ex} ,
\ee
\be
\label{oml}
\tilde{\Omega}'_{X}=
\tilde{\Omega}_{X} Z\left[-3\left(1+w_X ^{\rm
eff}\right)+2Z^2 \left(q+\frac{1}{Z^2}\right)\right]
\hspace{1ex} ,
\ee
where
\be
\label{qequ}
q=\frac{-\ddot{a}}{H^2a}=-\left(\frac{\dot{H}}{H^2}+1\right)=\frac{3\left(\tilde{\Omega}_{X}
w_X ^{\rm eff}+\tilde{\Omega}_{m}w_m ^{\rm
eff}\right)}{2Z^2}+\frac{1}{2Z^2}
\hspace{1ex}.
\ee
The evolution equation for $D$ is not coupled to the rest of the equations
(\ref{zeq}-\hspace{-0.1ex}\ref{oml}), so we
will not consider it for our dynamical study. Therefore, we will only
study the dynamical system for the variables
${\bf\tilde{\Omega}}\equiv(Z, \tilde{\Omega}_{m},
\tilde{\Omega}_{X})$, determined by the equations
(\ref{zeq}-\hspace{-0.1ex}\ref{oml}). The behavior of this system of
equations in the neighborhood of its stationary point is
determined by the corresponding matrix of its linearization. The
real parts of its eigenvalues tell us whether the corresponding
cosmological solution is stable or unstable with respect to the
homogeneous perturbations \cite{shta}.
%%%%%%%%%%%%%%%%%%%%%%%%%%%%%%%%%%%%%%%%%
\par\noindent
 To begin with, we have to find the critical points of this
 dynamical system, which can be written in vector form as follows
\begin{equation}
 {\bf\tilde{\Omega}'} ={\bf f}({\bf\tilde{\Omega}})
\end{equation}
where $f$ can be extracted from equations (\ref{zeq}-\hspace{-0.1ex}\ref{oml}). The
critical points, $\bf\tilde{\Omega}^\ast$, namely the points at which the
system will stay provided it is initially at there, are given by the
condition
\begin{equation}\label{con}
 f(\bf\tilde{\Omega}^\ast)=0\hspace{1ex}.
\end{equation}
Their dynamical character is determined by the eigenvalues of the
matrix
\begin{equation}
 \frac{\partial {\bf f}}{\partial\bf\tilde{\Omega}}\Bigg|_{\bf\tilde{\Omega}=\bf\tilde{\Omega}^\ast} .
\end{equation}
The coordinates of the critical points in the state space,
i.e. $\mathbf{\tilde{\Omega}}$ $=$ $(Z,\tilde{\Omega}_{m},\tilde{\Omega}_{X})$,
as well as their eigenvalues are given in the following table\\
\begin{center}
\begin{tabular}{ccc}
Model & Coordinates & Eigenvalues \\
\hline\\
$F_{-}$&$(-1,1,0)$&$-(1+3w_{m}^{\rm eff},3w_{m}^{\rm eff},3(w_{m}^{\rm eff}-w_{X}^{\rm eff}))$\\
$dS_{-}$&$(-1,0,1)$&$-(1+3w_{X}^{\rm eff},3(w_{X}^{\rm eff}-w_{m}^{\rm eff}),3w_{X}^{\rm eff})$\\
$E$&$(0,-\frac{(1+3w_{X}^{\rm eff})}{3(w_{m}^{\rm eff}-w_{X}^{\rm eff})},\frac{(1+3w_{m}^{\rm eff})}{3(w_{m}^{\rm eff}-w_{X}^{\rm eff})})$& $\alpha(0,-1,1)$\\
$F_{+}$&$(1,1,0)$&$(3w_{m}^{\rm eff},1+3w_{m}^{\rm eff},3(w_{m}^{\rm eff}-w_{X}^{\rm eff}))$\\
$dS_{+}$&$(1,0,1)$&$(3w_{X}^{\rm eff},3(w_{X}^{\rm eff}-w_{m}^{\rm eff}),1+3w_{X}^{\rm eff})$\\
\end{tabular}
\end{center}
with
\be
\alpha = \sqrt{\frac{-(1+3w_{m}^{\rm eff})(1+3w_{X}^{\rm eff})}{2}}\\
\hspace{1ex}.
\ee
The cosmological models identified with $F_{+}$ and $F_{-}$ are the expanding ($H>0$) and contracting ($H<0$) FLRW models ($k=\Lambda=0$) respectively,
$dS_{+}$ and $dS_{-}$ are the expanding and contracting de Sitter models ($k=\rho_{m}=0$) respectively,
and $E$ stands for the Einstein universe ($k=1$ and $H=0$).
\par\noindent
Concerning the dynamical character of the critical points, if the real part of
the eigenvalues of a critical point is not zero, the point is said
to be {\em hyperbolic} \cite{CS}. In this case, the dynamical
character of the critical point is determined by the sign of the
real part of the eigenvalues:  if all of them are positive, the
point is said to be a {\em repeller}, because arbitrarily small
deviations from this point will move the system away from this
state.  If all of them are negative the point is called an {\em
attractor} because if we move the system slightly from this point in
an arbitrary way, it will return to it. Otherwise, we say the
critical point is a {\em saddle} point.
As we can see from the previous table the dynamical character of all equilibrium points
depend on the equation of state parameters, i.e. $w_{m}^{\rm eff}$ and $w_{X}^{\rm eff}$.

\begin{center}
\begin{tabular}{ccc}
Model &\hspace{10ex}Repeller\hspace{10ex}&\hspace{10ex}Attractor\hspace{10ex} \\
\hline\\
$F_{-}$&\hspace{10ex}$w_{m}^{\rm eff}<-\frac{1}{3}$ and $w_{m}^{\rm eff}<w_{X}^{\rm eff}$ &\hspace{10ex}- \\
$dS_{-}$&\hspace{10ex}$w_{X}^{\rm eff}<-\frac{1}{3}$ and $w_{m}^{\rm eff}>w_{X}^{\rm eff}$ & \hspace{10ex}-\\
$E$&\hspace{10ex}-&\hspace{10ex}-\\
$F_{+}$&\hspace{10ex}-& $w_{m}^{\rm eff}<-\frac{1}{3}$ and $w_{m}^{\rm eff}<w_{X}^{\rm eff}$\\
$dS_{+}$&\hspace{10ex}-& $w_{X}^{\rm eff}<-\frac{1}{3}$ and $w_{m}^{\rm eff}>w_{X}^{\rm eff}$\\
\end{tabular}
\end{center}

\par\noindent
It should be pointed out that for the Einstein universe ($E$-model) there is no way to be a repeller or, an attractor.
It represents a set of infinite saddle points which have to satisfy one of the following conditions
\ba
w_{m}^{\rm eff}>-\frac{1}{3}\hspace{2ex}  \mbox{and} \hspace{2ex}  w_{X}^{\rm eff}<-\frac{1}{3} \\
\mbox{or},\hspace{3ex} w_{m}^{\rm eff}<-\frac{1}{3}\hspace{2ex}
\mbox{and}\hspace{2ex}  w_{X}^{\rm eff}>-\frac{1}{3}
\hspace{1ex}.
\ea
%%%%%%%%%%%%%%%%%%%%%%%%%%%%%%%%%%%%%%%%%%%%%%%%%%%%%%%%%%%%%%%%%%%%%%%%%%%%%%%
%%%%%%%%%%%%%%%%%%%%%%%%%%%%%%%%%%%%%%%%%%%%%%%%%%%%%%%%%%%%%%%%%%%%%%%%%%%%%%%
\section{Conclusions}
Understanding the dark energy is one of the biggest challenges to
the particle physics this century. Studying the interaction between
the dark energy and ordinary matter will open a possibility of
detecting the dark energy. Being a dynamical component, the scalar
field dark energy is expected to interact with the ordinary matters.
For example, Carroll \cite{ca} (see also \cite{ming}) has considered
an interaction of form $Q F_{\mu\nu}\tilde{F^{\mu\nu}}$ with $
F_{\mu\nu}$ being the electromagnetic field strength tensor which
has interesting implication on the rotation of the plane of
polarization of light coming from distant sources. Recent data on
the possible variation of the electromagnetic fine structure
constant has triggered interests in studies related to the
interactions between
quintessence and the matter fields.\\
In this paper we have studied the cosmological dynamics of the
Interacting Holographic Dark Energy model. We derived the system of
the first-order differential equations which describes the evolution
of the three dimensionless quantities that form a compact state
space. Furthermore, the five critical points of the aforementioned
cosmological model were obtained and the dynamical character of
these critical points was presented.
\par\noindent
In particular, it was shown that $E$-model
represents a set of infinite saddle points whose line element is that of
the Einstein universe.
By using the fact that the equation of state parameters $w_{m}^{eff}$ and
$w_{X}^{eff}$ should always be negative as indicated by equations (\ref{we1}) and (\ref{we2}),
we obtained that if $\left|3w_{m}^{eff}\right|<1$ the critical points of the $F_{\pm}$
models are  saddle points. On the other hand, if $\left|3w_{m}^{eff}\right|>1$,
the critical point of model $F_{-}$ will be a repeller under the condition $|w_{m}^{eff}|>| w_{X}^{\rm eff}|$,
while the critical point of the $F_{+}$ model will be an attractor.
Concerning the $dS_{\pm}$ models, we see that if $|3w_{X}^{eff}|<1$ then the
critical points of $dS_{\pm}$  are saddle points. On the other
hand, if $|3w_{X}^{eff}|>1$, the critical point of model $dS_{-}$ model
will be a repeller under the condition $|w_{m}^{eff}|<| w_{X}^{\rm eff}|$, while
the critical point of $dS_{+}$ will be an attractor.
\par\noindent
Finally, it should be stressed that evidence was recently provided
by the Abell Cluster A586 in support of the interaction between
dark energy and dark matter \cite{Bertolami:2007zm}. However,
despite the fact that numerous works have been performed till now,
there are no strong observational bounds on the strength of this
interaction  \cite{Amendola:2006dg}.
This weakness to set stringent (observational or theoretical)
constraints on the strength of the coupling between dark energy and dark
matter stems from our unawareness of the nature and origin of dark components of the Universe.
It is therefore more than obvious that further work is needed to this direction.
%%%%%%%%%%%%%%%%%%%%%%%%%%%%%%%%%%%%%%%%%%%%%%%%%%%%%%%%%%%%%%%%%%%%%%%%%%%%%%%
%%%%%%%%%%%%%%%%%%%%%%%%%%%%%%%%%%%%%%%%%%%%%%%%%%%%%%%%%%%%%%%%%%%%%%%%%%%%%%%
\section*{Acknowledgements}
ECV is supported by the Greek State Scholarship Foundation (I.K.Y.).
%%%%%%%%%%%%%%%%%%%%%%%%%%%%%%%%%%%%%%%%%%%%%%%%%%%%%%%%%%%%%%%%%%%%%%%%%%%%%%%%
%%%%%%%%%%%%%%%%%%%%%%%%%%%%%%%%%%%%%%%%%%%%%%%%%%%%%%%%%%%%%%%%%%%%%%%%%%%%%%%%

\end{document}